# buckling of thin composite plates reinforced with randomly oriented, straight single-walled carbon nanotubes using B3-Spline finite strip method

H. Foroughi[a], H.R. Askariyeh[b] and M. Azhari[a, 1]

[a]Department of Civil Engineering, Isfahan University of Technology, Isfahan 84156-83111, Iran
[b]Department of Civil Engineering, Faculty of Engineering, Yazd University, Yazd, P.O.BOX 89195-741, Iran

**Abstract**

This paper is devoted to the buckling analysis of thin composite plates under straight single-walled carbon nanotubes reinforcement with uniform distribution and random orientations. In order to develop the fundamental equations, the B3-Spline finite strip method along with the classical plate theory (CPT) is employed and the total potential energy is minimized which leads to an eigenvalue problem. For deriving the effective modulus of thin composite plates reinforced with carbon nanotubes, the Mori-Tanaka method is used in which each straight carbon nanotube is modeled as a fiber with transversely isotropic elastic properties. The numerical results including the critical buckling loads for rectangular thin composite plates reinforced by carbon nanotubes with various boundary conditions and different volume fractions of nanotubes are provided and the positive effect of using carbon nanotubes reinforcement in buckling of thin plates is illustrated.

[1]**Corresponding Author:**

Mojtaba Azhari, Department of Civil Engineering, Isfahan University of Technology, Isfahan, Iran.

Phone: +98 3113913804;  Fax: ++98 3113912700;  Email: mojtaba@cc.iut.ac.ir

## 1.Introduction

In recent years, there has been a growing trend towards using composite materials in many engineering fields. Many researchers have attempted to reinforce composite structures by high strength reinforcements, among which the carbon nanotubes are widely used.

In this regard, many researchers have elaborated on the material properties of the composites reinforced with CNTs. Among these investigations, it can be referred to Fukuda and Kawatawho evaluated the Young modulus of composite structures reinforced by short fibers [1]. Also Ajayan et al. studied the polymer composites reinforced by aligned CNT arrays [2]. Griebel and Hamaekersused the molecular dynamics simulations to evaluate the elastic moduli of CNT reinforced composite structures [3]. Esawi and Faragshowed that reinforcing composite structures by CNTs leads to a considerably high stiffness while economical [4].

In order to examine the behavior of CNT reinforced composites in bending and buckling, several researches have been conducted. Vodenitcharova and Zhang studied the bending and local buckling of a nanocomposite beam reinforced by a single-walled CNT by employing the Airy stress-function method to analyze the deformation of the matrix [5]. Shen studied the postbuckling of nanotube-reinforced composite cylindrical shells in thermal environments [6]. Zhu et. Al discussed the static and free vibration of composite plates under CNT reinforcement by employing the finite element method [7].

Finite strip method is a modified finite element method introduced by Cheung and has been obtained by discretizing the plate into some finite strips [8]. Bradford and Azhari investigated the buckling of plates with different end conditions using the finite strip method [9].Azhari et al. applied the spline finite strip method augmented by the bubble functions for elastic analysis of rectangular plates to show that efficiently improves the convergence of the spline finite strip method with respect to strip subdivision [10]. Lotfi et al. studied the inelastic local buckling of skew thin thickness-tapered plates using Spline finite strip method [11].

In the current study, the buckling analysis of thin composite plates reinforced with single-walled CNTs is conducted. The distribution of nanotubes in the matrix is uniform and they are randomly oriented. In order to derive the fundamental equations to be applied in the buckling relations, the classical plate theory and B3-Spline finite strip method is used. It is noteworthy that the characterization of the CNTs properties is obtained through the Mori-Tanaka approach.

## 2. Eshelby-Mori-Tanaka approach

The Mori-Tanaka approach is introduced as an analytical method to characterize a linear elastic polymer matrix reinforced by a large number of dispersed CNTs that are aligned or randomly oriented, straight and of infinite length.
This method assumes that each nanotube is embedded in an infinite matrix subjected to an effective stress ($\sigma_m$) or an effective strain ($\varepsilon_m$).According to Benveniste the effective elastic tensor is revised as follows [12]

$$\mathbf{C} = \mathbf{C}_m + V_{CN}((\mathbf{C}_{CN} - \mathbf{C}_m).\mathbf{A}).[V_m \mathbf{I} + V_{CN} \mathbf{A}]^{-1} \tag{1}$$

where $\mathbf{I}$ is the fourth-order identity tensor, the subscripts $m$ and $CN$ denote the quantities of the matrix and the nanotubes, respectively, $V_m$ and $V_{CN}$ stand for the volume fractions, and $\mathbf{C}_m$ and $\mathbf{C}_{CN}$ denote the elastic moduli tensors and $\mathbf{A}$ is the fourth-order tensor depending on the characteristics of the carbon nanotube.
in which

$$\mathbf{A} = [\mathbf{I} + \mathbf{S}.(\mathbf{C}_m)^{-1}.(\mathbf{C}_{CN} - \mathbf{C}_m)]^{-1} \tag{2}$$

where $\mathbf{S}$ is the Eshelby tensor derived by Mura [13].
In this study, the CNTs are assumed to be randomly oriented in the matrix. The orientation of a straight CNT is described by two Euler angles $\alpha$ and $\beta$. In order to transform in global $(x,y,z)$ coordinates to local $(x',y',z')$ coordinates, the transformation matrix $\mathbf{g}$ is used as follows

$$\mathbf{g} = \begin{bmatrix} \cos\beta & -\cos\alpha\sin\beta & \sin\alpha\sin\beta \\ \sin\beta & \cos\alpha\cos\beta & -\sin\alpha\cos\beta \\ 0 & \sin\alpha & \cos\alpha \end{bmatrix} \tag{3}$$

Moreover, the orientation distribution of CNTs in a composite is characterized by a probability density function $\rho(\alpha,\beta)$ satisfying the normalization condition as below

$$\int_0^{2\pi} \int_0^{\frac{\pi}{2}} \rho(\alpha,\beta) \sin\alpha \, d\alpha \, d\beta = 1 \tag{4}$$

In the above relation, in case that the orientations of CNTs are completely random, the density function is defined as $\rho(\alpha,\beta) = \pi/2$.
Considering the constitutive equations of the matrix phase, the strain and the stress vectors $\boldsymbol{\varepsilon}_{CN}(\alpha,\beta)$ and $\boldsymbol{\sigma}_{CN}(\alpha,\beta)$ are related to the stress matrix $\boldsymbol{\sigma}_m$ as below

$$\boldsymbol{\varepsilon}_{CN}(\alpha,\beta) = \mathbf{A}(\alpha,\beta).\mathbf{C}_m^{-1}.\boldsymbol{\sigma}_m, \quad \boldsymbol{\sigma}_{CN}(\alpha,\beta) = [\mathbf{C}_r.\mathbf{A}(\alpha,\beta).\mathbf{C}_m^{-1}].\boldsymbol{\sigma}_m \tag{5}$$

The average strain and stress in all randomly oriented CNTs can be written as

$$\boldsymbol{\varepsilon}_{CN}^{avg} = \left[\int_0^{2\pi} \int_0^{\frac{\pi}{2}} \rho(\alpha,\beta) \mathbf{A}(\alpha,\beta) \sin\alpha \, d\alpha \, d\beta\right].\boldsymbol{\varepsilon}_m,$$

$$\boldsymbol{\sigma}_{CN}^{avg} = \left[\int_0^{2\pi} \int_0^{\frac{\pi}{2}} \rho(\alpha,\beta)[\mathbf{C}_{CN}.\mathbf{A}(\alpha,\beta).\mathbf{C}_m^{-1}] \sin\alpha \, d\alpha \, d\beta\right].\boldsymbol{\sigma}_m \tag{6}$$

where, the superscript "$avg$" denotes the average over special orientations. According the average theorems and substituting Eq. (9) in to the constitutive relation $\boldsymbol{\sigma} = \mathbf{C}.\boldsymbol{\varepsilon}$, the effective elastic modulus tensor for randomly oriented CNTs in the matrix will be derived, which is not given here for brevity.
According to the Mori-Tanaka approach, the composite is considered isotropic, and so its bulk modulus $K$ and shear modulus $G$ are defined as

$$K = K_m + \frac{V_{CN}(\delta_{CN} - 3K_m \alpha_{CN})}{3(V_m + V_{CN} \alpha_{CN})}, \quad G = G_m + \frac{V_{CN}(\eta_{CN} - 2G_m \beta_{CN})}{2(V_m + V_{CN} \beta_{CN})} \tag{7}$$

where $K_m$ and $G_m$ are the bulk and shear moduli of the matrix phase, respectively, and

$$\alpha_{CN} = \frac{3(K_m + G_m) + k_{CN} - l_{CN}}{3(G_m + k_{CN})},$$

$$\beta_{CN} = \frac{1}{5}\left[\frac{4G_m + 2k_{CN} + l_{CN}}{3(G_m + k_{CN})} + \frac{4G_m}{G_m + p_{CN}} + \frac{2[G_m(3K_m + G_m) + G_m(3K_m + 7G_m)]}{G_m(3K_m + G_m) + m_{CN}(3K_m + 7G_m)}\right],$$

$$\delta_{CN} = \frac{1}{3}\left[n_{CN} + 2l_{CN} + \frac{(2k_{CN} + l_{CN})(3K_m + 2G_m - l_{CN})}{G_m + k_{CN}}\right],$$

$$\eta_{CN} = \frac{1}{5}\left[\frac{2}{3}(n_{CN} - l_{CN}) + \frac{8G_m p_{CN}}{G_m + p_{CN}} + \frac{8m_{CN} G_m(3K_m + 4G_m)}{3K_m(m_{CN} + G_m) + G_m(7m_{CN} + G_m)} + \frac{2(k_{CN} - l_{CN})(2G_m + l_{CN})}{3(G_m + k_{CN})}\right] \quad (8)$$

where $k_{CN}$, $l_{CN}$, $m_{CN}$, $n_{CN}$ and $p_{CN}$ are taken from analytical results of Popov et. al (2000). Finally, the effective Young's modulus $E$ and Poisson's ratio $\upsilon$ of the composite in terms of the bulk modulus $K$ and shear modulus $G$ is defined as

$$E = \frac{9KG}{3K + G}, \quad \upsilon = \frac{3K - 2G}{6K + 2G} \quad (9)$$

The constitutive stress-strain relations are then introduced as

$$\begin{Bmatrix} \sigma_{xx} \\ \sigma_{yy} \\ \sigma_{xy} \end{Bmatrix} = \begin{bmatrix} Q_{11} & Q_{12} & 0 \\ Q_{12} & Q_{22} & 0 \\ 0 & 0 & Q_{66} \end{bmatrix} \begin{Bmatrix} \varepsilon_{xx} \\ \varepsilon_{yy} \\ \varepsilon_{xy} \end{Bmatrix}, \quad Q_{11} = Q_{22} = \frac{E(z)}{1-\nu^2}, Q_{12} = \nu Q_{11}, Q_{66} = G \quad (10)$$

## 3. Mechanical buckling

### 3.1. The displacement field

In this section, mechanical buckling of thin composite plates reinforced with single-walled CNTs is discussed. Based on the classical plate theory (CPT), the displacement field is defined as

$$u(x, y, z) = u_0(x, y) - z\frac{\partial w_0}{\partial x} \quad v(x, y, z) = v_0(x, y) - z\frac{\partial w_0}{\partial y} \quad w(x, y, z) = w_0(x, y) \quad (11)$$

where $u_0$, $v_0$ and $w_0$ are the mid-plane displacements.

### 3.2. B3-Spline Finite strip approach

According to the classical plate theory for analysis thin plates, the ensuring displacement fields for one strip is

$$\delta_n = \left\{w_i, \left(\frac{\partial w}{\partial x}\right)_i, w_j, \left(\frac{\partial w}{\partial x}\right)_j\right\}_n^T = \begin{Bmatrix} w_{in} \\ \theta_{in} \\ w_{jn} \\ \theta_{jn} \end{Bmatrix} = \begin{Bmatrix} \boldsymbol{\delta}_{in} \\ \boldsymbol{\delta}_{jn} \end{Bmatrix} \quad (12)$$

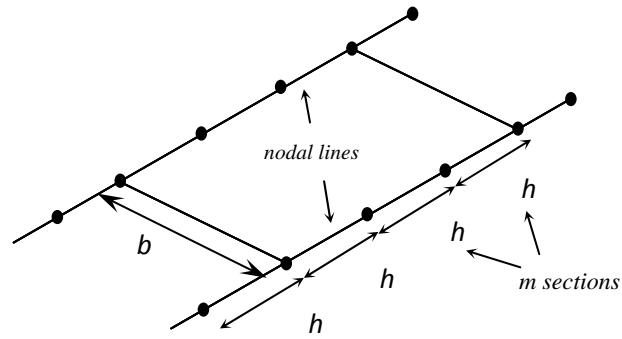

*Fig. 1.nodal lines for one strip in Spline finite strip method*

A generic function $f(y)$, say a generalized, can be represented as a linear combination of local B3- Spline functions $\psi_i(y)$ and parameters $\alpha_i$ for the *j*th node, according to the expression:

$$f(y) = \sum_{i=-1}^{m+1} \alpha_i \psi_i(y) \tag{13}$$

where

$$\psi_i(y) = \left(\frac{1}{6h^3}\right)\begin{cases} (y-y_{i-2})^3 & y_{i-2} \leq y \leq y_{i-1} \\ h^3 + 3h^2(y-y_{i-1}) + 3h(y-y_{i-1})^2 - 3(y-y_{i-1})^3 & y_{i-2} \leq y \leq y_{i-1}, \\ h^3 + 3h^2(y_{i+1}-y) + 3h(y_{i+1}-y)^2 - 3(y_{i+1}-y)^3 & y \leq y \leq y_{i+1}, \\ (y_{i+2}-y) & y_{i+1} \leq y \leq y_{i+2}, \end{cases} \tag{14}$$

The shape function, $w_{ij}$ could be introduced as

$$w_{ij} = N_i(x)\psi_j(y) \tag{15}$$

In addition, $N_i(x)$ are the following Hermitian cubic polynomials.

$$N_1(x) = 1 - 3\left(\frac{x}{b}\right)^2 + 2\left(\frac{x}{b}\right)^3, \; N_2(x) = x\left[1 - 2\left(\frac{x}{b}\right) + \left(\frac{x}{b}\right)^2\right], \; N_3(x) = 3\left(\frac{x}{b}\right)^2 - 2\left(\frac{x}{b}\right)^3, \; N_4(x) = x\left[\left(\frac{x}{b}\right)^2 - \left(\frac{x}{b}\right)\right] \tag{16}$$

The out of plane buckling displacement of the plate, $(w)$ can be expressed as

$$w(x,y) = \sum_{i=1}^{4} \sum_{j=-1}^{m+1} \left(N_i(x).\psi_j(y).\delta_{ij}\right) \tag{17}$$

$$\{\delta\} = \{w_{1,-1}, w_{1,0}, w_{1,1}, ..., w_{1,m+1}, \theta_{1,-1}, \theta_{1,0}, \theta_{1,1}, ..., \theta_{1,m+1}, w_{1,-1}, w_{1,0}, w_{1,1}, ..., w_{1,m+1}, \theta_{2,-1}, \theta_{2,0}, \theta_{2,1}, ..., \theta_{2,m+1}\}^T \tag{18}$$

In Eq.(19) and Eq.(20), $\{\delta\}$, is the nodal displacement vector. Furthermore, the following strain- displacement relations are employed.

$$\varepsilon_{ij} = \frac{1}{2}\left(u_{i,j} + u_{j,i} + u_{k,i}u_{k,j}\right) \tag{19}$$

Assuming that the plate is thin based on Kirchhorf hypothesis [14], the vectors of internal moments, and infinitesimal buckling curvatures, may be given by well-known expressions:

$$\{\sigma_f\} = \{M_x, M_y, M_{xy}\}^T, \{\varepsilon_f\} = \{\rho_x, \rho_y, \rho_{xy}\}^T \tag{20}$$

In which

$$\rho_x = \frac{\partial^2 w}{\partial x^2}, \rho_y = \frac{\partial^2 w}{\partial y^2}, \rho_{xy} = \frac{\partial^2 w}{\partial x \partial y} \tag{21}$$

## 4. The derivation of the critical buckling load

On the basis of the present approach and according to the total potential theorem, the total potential energy of a plate can be expressed as

$$U_{total} = U + V_p \tag{22}$$

in which $U$ is the strain energy of the plate and $V_p$ is potential energy of the external loads.

According the procedure of minimizing the total potential energy in the standard finite strip method, the strain energy and stiffness matrix can obtain as

$$U = \frac{1}{2}\iint_\Omega \varepsilon_f^T \sigma_f \, dA = \frac{1}{2}\sum\sum \delta^T \mathbf{K} \delta, \quad \mathbf{K} = \iint_\Omega [\mathbf{B}_f]^T [\mathbf{Q}][\mathbf{B}_f] \, dA \tag{23}$$

in which

$$[B_f] = \begin{bmatrix} \dfrac{-\partial^2 ([N][\psi])}{\partial x^2} \\ \dfrac{-\partial^2 ([N][\psi])}{\partial y^2} \\ \dfrac{-\partial^2 ([N][\psi])}{\partial x \partial y} \end{bmatrix}_{3 \times c} \tag{24}$$

where $\mathbf{K}$ is the stiffness matrix of the plate.

Following the procedure typically used in B3-Spline finite strip method, the geometry stiffness matrix of the strip for a plate subjected to $\sigma_x^0$, $\sigma_y^0$ and $\sigma_{xy}^0$ in-plane stresses

$$V_p = \frac{1}{2}\iiint \left\{\sigma_x^0\left(\frac{\partial w}{\partial x}\right)^2 + \sigma_y^0\left(\frac{\partial w}{\partial y}\right)^2 + 2\sigma_{xy}^0 \frac{\partial w}{\partial x}\frac{\partial w}{\partial y}\right\} dV = \frac{1}{2}\sum\sum \delta^T \mathbf{K}_g \delta \tag{25}$$

where $\mathbf{K}_g$ is the geometric stiffness matrix and may be obtained as

$$[K_g] = t \iint_\Omega [B_g]^T [\sigma_g][B_g] \tag{26}$$

In which

$$[B_g] = \begin{bmatrix} \dfrac{\partial([N][\psi])}{\partial x} \\ \dfrac{\partial([N][\psi])}{\partial y} \end{bmatrix}_{2\times c} \quad , \quad [\sigma_g] = \begin{bmatrix} \sigma_x^0 & \sigma_{xy}^0 \\ \sigma_{xy}^0 & \sigma_y^0 \end{bmatrix} \tag{27}$$

All of the equations are developed for one strip. After assembling the stiffness matrix of the plate, and the geometric stiffness matrix over all the strips of the plate, followed by imposing the boundary conditions. The critical buckling load of a composite plate reinforced by CNTs, defined as $\sigma_{cr}$, can be determined by solving the following eigenvalue problem using standard computer subroutines.

$$|\mathbf{K} + \sigma_{cr}\mathbf{K}_g| = 0 \tag{28}$$

## 4. Numerical results

In this section, several numerical examples are presented and discussed to illustrate the efficiency of the B3-Spline finite strip method in mechanical buckling analysis of thin rectangular plates reinforced with single-walled CNTs. The matrix in which the CNTs are embedded is PmPV (Poly{(m-phenylenevinylene)-co-[(2.5-dioctoxy-p-phenylene)vinylene]}, the material properties of which are $v_m = 0.34$ and $E_m = 2.1\ GPa$ at room temperature of $T = 300K$.

In order to validate the results obtained by the present method, the normalized critical buckling load ($\lambda$) for composite plates with no CNT ($V_{CN} = 0$) are presented in Table 1 and compared with those obtained by Bradford and Azhari [13]. In Table 1, the results are provided for various boundary conditions. As it can be observed from Table 1, the results obtained from the present method have excellent accuracy.

*Table 2. Normalized critical buckling factor ( $\lambda = \dfrac{(\sigma_{cr} \times 12(1-v^2) \times b^2)}{(\pi^2 E h^3)}$ ) for different boundary conditions*

| $b/h$ | Methods | Boundary conditions | | | |
|---|---|---|---|---|---|
| | | SSSS | SCSC | SCSS | CCCC |
| 0.01 | *present* | 4.000 | 7.721 | 5.979 | 10.072 |
| | *Bradford et al[13]* | 4.000 | 7.72 | 5.97 | 10.08 |

Table 3 illustrates the values of $\lambda$ for CNT reinforced composite plates of Fig. 1 for different nanotube volumes ($V_{CN}$). Here again the results for various boundary are discussed. As it was expected, by increasing the volume fraction of CNTs, the normalized critical buckling load increases as a result. Moreover, by comparing the results for different boundary conditions, it can be seen that the normalized critical buckling load for the free boundary conditions, showed with F in the tables, are considerably lower than those for the clamped boundary conditions which is quite realistic. Furthermore, as the volume fraction of CNTs increases, the normalized critical buckling loads increases which shows the influence of CNTs in increasing the stiffness of the plate.

*Table 3. Normalized critical buckling loads ( $\lambda = \frac{(\sigma_{cr} \times 12(1-\upsilon_m^2) \times b^2)}{(\pi^2 E_m h^3)}$ ) for various carbon nanotubes volume fraction and different boundary conditions*

| $V_{CN}$ | $b/h$ | SCSC | SCSS | SSSS | SFSF |
|---|---|---|---|---|---|
| 0.01 | 100 | 10.512 | 7.768 | 5.390 | 1.271 |
| | 50 | 10.320 | 7.711 | 5.375 | 1.269 |
| 0.05 | 100 | 21.351 | 15.774 | 10.944 | 2.607 |
| | 50 | 20.960 | 15.658 | 10.913 | 2.604 |
| 0.1 | 100 | 35.167 | 25.978 | 18.023 | 2.607 |
| | 50 | 34.522 | 25.787 | 17.972 | 2.604 |

In Fig 2, the critical buckling load ($N_{cr}$) versus aspect ratio ($a/b$) for composite plates reinforced with three CNTs volume fractions of 0.01, 0.05 and 0.1 is depicted. It is observed that the normalized critical buckling load increases with the volume fraction of CNTs. Moreover, as proved in the previous studies, by the increase in the aspect ratio, the critical buckling coefficient converges.

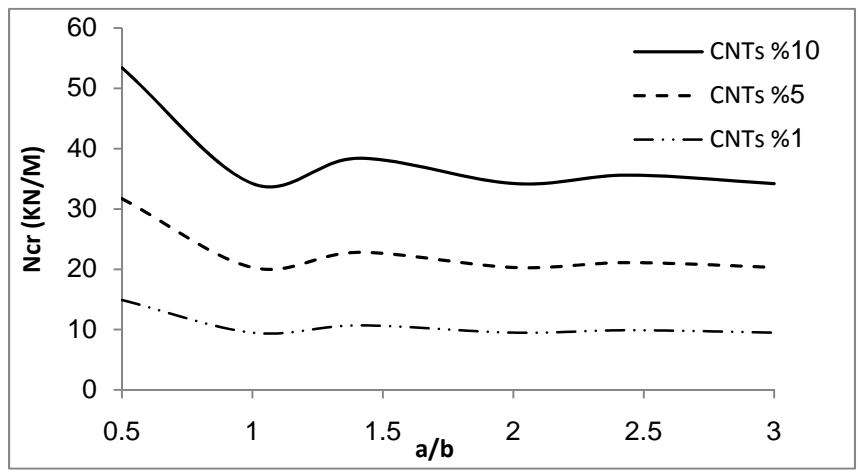

*Fig 2. Critical buckling load of composite plate reinforced with various volume fraction of carbon nanotubes versus aspect ratio*

## 5. Conclusions

In this paper, the B3-Spline finite strip method is successfully applied to the mechanical buckling analysis of thin composite plates reinforced with single-walled carbon nanotubes. The mechanical properties of carbon nanotubes embedded in the matrix are derived through the Mori-Tanaka approach. Comparison studies were conducted on the critical buckling load of composites without nanotubes to verify the present method and the results were found to be in good agreement with those available in literature. Several numerical examples for composites under carbon nanotubes reinforcement were provided and it was concluded that the change in carbon nanotubes volume fraction, boundary conditions, aspect ratio, thickness to width strongly affect the critical buckling load of the plate.

# References

[1] Fukuda, H., Kawata, K. (1974). "On Young's modulus of short fibre composites."*Fibre.Sci.Technol.,*7, 207-222.

[2] Ajayan, P.M., Stephan, O., Colliex, C., and Trauth, D. (1994)."Aligned carbon nanotube arrays formed by cutting a polymer resin—nanotube composite."*Science.,*265(5176), 1212-1214.

[3] Griebel, M., Hamaekers, J. (2004) "Molecular dynamics simulations of the elastic moduli of polymer–carbon nanotube composites."*Mech. Eng.,* 193(17-20), 1773-1788.

[4] Esawi , M. K., Farag, M. (2007). "Carbon Nanotube Reinforced Composites: Potential and Current Challenges." *Mater.Des.,*28(9), 2394-2401.

[5] Vodenitcharova, T., Zhang, L. C. (2006)."Bending and local buckling of a nanocomposite beam reinforced by a single-walled carbon nanotube."*Int. J. Solids. Struct.,*43(10), 3006-3024.

[6] Shen, H. S. (2011). "Postbuckling of nanotube-reinforced composite cylindrical shells inthermal environments – Part I: Axially-loaded shells."*Compos.Struct.,* 93(8), 2096-2108

[7] Zhu, P., Lei, Z.X., Liew, K.M. (2012). "Static and free vibration analyses of carbon nanotubereinforced composite plates using finite element method with first order shear deformation plate theory." *Compos.Struct.,* 94(4), 1450-60.

[8] Cheung, Y.K. (1976). *"Finite strip method in structural analysis."* Pergamon Press, Oxford-New York.

[9] Bradford, M.A., Azhari, M. (1995). "Buckling of plates with different end conditions using the finite strip method." *Comput.Struct.,* 56(1), 75-83.

[10]Azhari, M., Hoshdar, S., Bradford, M.A. (2000)."On the use of bubble functions in the local buckling analysis of plate structures by the spline finite strip method."*Int J Numer Methods Eng*., 48, 583-593.

[11] .Lotfi , S., Azhari , M., Heidarpour , A. (2011). "Inelastic initial local buckling of skew thin thickness-tapered plates with and without intermediate supports using the isoparametric spline finite strip method."*Thin-Walled Structures.,* 49, 1475-1482.

[12] Benveniste, Y. (1987). "A new approach to the application of Mori–Tanaka's theory in composite materials."*Mech Mater.,* 6(2), 147-157.

[13] Mura, T.(1987). "*Micromechanics of Defects in Solids.*"MartinusNijhoff Publishers, Dordrecht.

[14] Vrcelj, Z., Bradford, M.A. (2008). "A simple method for the inclusion of external and internal supports in the spline finite strip method (SFSM) of buckling analysis."*ComputStruct*., 86, 529-544.

[15] Foroughi, H., Askariyeh, H.R., Azhari, M. (2013).'Mechanical Buckling of Thick Composite Plates Reinforced with Randomly Oriented, Straight, Single-Walled Carbon Nanotubes Resting on an Elastic Foundation using the Finite Strip Method." *Journal of Nanomechanics and Micromechanics.,* 3, 49-58.